\documentstyle[11pt,newpasp,twoside,epsf]{article}
\begin{document}

\title{Radiative transfer in 3D}

\author{M. Juvela}
\affil{Helsinki University Observatory, FIN-00014 University of Helsinki,
Finland}
\author{P. Padoan}
\affil{Harvard University Department of Astronomy}

\begin{abstract}
The high resolution and sensitivity provided by ALMA will reveal new small
scale structures in many sources, e.g. star forming regions. Such
inhomogeneities may not have been considered in the analysis of past
observations but they will be essential to the understanding of future data.
Radiative transfer methods are needed to interpret the observations and the
presence of complicated source structures and small scale inhomogeneities
requires 3D modelling.

We will describe studies of molecular line emission we have made using
models of inhomogeneous molecular clouds. 
These are based on MHD models and include e.g. thermal balance calculations
for molecular clouds.

Our radiative transfer code is based on Monte Carlo simulation. With current
computing resources it has already become possible to handle even large 3-D
models. We shall show that Monte Carlo method can be applied also in cases
of hight optical depths.

\end{abstract}


\section{Introduction}

Interpretation of molecular line observations requires understanding of the
radiative transfer processes involved. The observed intensity is an integral
over the line-of-sight and depends on various physical parameters that vary
along this line. The problem is usually a non-invertible one. Finite angular
resolution means that conditions may vary over the beam and further
complications arise e.g. in the form of incomplete beam filling.

One may approach the problem from different perspective by constructing
models of the density, velocity and temperature structure of the object.
Models can be tested by using radiative transfer methods that predict the
observed spectra. Eventually one can find a model that fits the
observations. However, usually there is a number of possible solutions.

The models can be restricted by other arguments e.g. by requiring physical
self-consistency, although e.g. in massive star forming regions the
situation is often too complicated for this (e.g. Juvela 1998). For
interstellar clouds at larger scales the situation seems to be better. There
the dynamical processes are less violent and the laws of
magneto-hydrodynamics can be used to simulate typical distribution and
motion of gas. Conversely, with radiative transfer methods the basic
assumptions of the MHD calculations can be checked against observations.

\section{Radiative transfer with the Monte Carlo method}

For radiative transfer calculations the model cloud is discretized into
cells e.g. according to a three-dimensional Cartesian grid. Each cell is
assigned density, temperature, velocity and intrinsic linewidth. The effect
of the radiation field is simulated and the results are used to update
estimates of the level populations in each cell. Final solution is obtained
by iterating these two steps.

There are several methods for improving the computational efficiency of
Monte Carlo. These include the use of a reference field (Bernes 1979; Choi
et al. 1995), importance sampling and the use of quasi random numbers
(Juvela 1997). Furthermore, if one uses the same set of random numbers on
each iteration one can eliminate random noise from computed level
populations. This enables the use of similar acceleration methods as used
with lambda iteration e.g. Ng-acceleration (Ng 1974).

The Monte Carlo method has been considered unsuitable for use with high optical
depths, $\tau>>1.0$. Hartstein \& Liseau (1998) showed, however, that with
core saturation method calculations are possible even with optical depths of
several thousands. The idea is to consider only photons in the line wings.
In the optically thick case photons in the line centre are emitted and
absorbed locally and do not contribute to the transfer of energy.

We have used two simulation methods in the radiative transfer calculations.
Method B in Juvela (1997) is based on ray-tracing. A photon package is
started at the edge of the cloud and as the package moves a distance $s$
through a cell with optical depth $\tau$ the number of emitted photons escaping
the cell,
\begin{equation}
n(\nu) \sim n_u A_{ul} \frac{1- e^{- \tau}}{\tau} \phi(\nu),
\end{equation}
is added to the package while the rest of the emitted photons are absorbed
in the cell (Juvela 1997). In the formula $n_{u}$ is the population of the
upper energy level and $A_{ul}$ is the Einstein coefficient of spontaneous
emission. We can use this information in a way similar to the core
saturation method to eliminate photons that are absorbed in the same cell
from which they were emitted. The equilibrium equations must also be
modified, which can be done in several ways. We store in each cell the
fraction of discarded photons and use those numbers to correct the
equilibrium equations. This means an increase in the memory consumption by
$\sim$30\%. Conceptually the method is identical to the accelerated lambda
iteration (Olson, Auer \& Buchler 1986).

In Monte Carlo simulation the knowledge of the cell geometry is required
only for finding the next cell boundary along the track of the photon
package and for calculating the cell volumes. This makes it easy design the
computational geometry according to the problem at hand. So far we have
implemented the following: (1) spherically symmetric clouds divided into
shells, (2) cylinder symmetric clouds divided by cylinders and orthogonal
planes, (3) 3D Cartesian grid with cubic cells, (4) 3D grid with embedded
spherical clumps divided by spheres, longitudes and latitudes and (5) 3D
Cartesian grid with hierarchical subdivision according chosen criteria. The
last option is promising for the MHD simulations where better resolution is
usually needed only in some small sub-volume. The reduced memory
requirements are, however, not complemented by equal savings in the run
times although importance sampling can be used to concentrate on the regions
with higher discretization.

\section{MHD models of interstellar cloud}

Padoan \& Nordlund (1999) have shown that super-alfv\'enic random flows
provide a good model for the dynamics and structure of molecular clouds. The
column density maps made of MHD simulations are reminiscent of the
filamentary structures observed in interstellar clouds and the results give
a good starting point for the radiative transfer modelling of these clouds.
With a snapshot of the MHD simulations (i.e. density and velocity fields)
the Monte Carlo method can be used to solve the the radiative transfer
problem and the computed spectra can be compared with observations.

The original MHD calculations were performed on 128$^3$ cell grid. For
radiative transfer calculations the data are resampled on a smaller grid of
typically 90$^3$ or 64$^3$ cells. This can be done without significant loss
of accuracy since individual cells are usually optically thin. However,
proper sampling of the velocity field sets its own requirements on the
discretization. Models represent clouds with linear size $\sim$10\,pc and
average gas densities a few times 100\,cm$^{-3}$. In the densest knots the
density exceeds 10$^4$\,cm$^{-3}$.

The results of the radiative transfer calculations are presented as
synthetic molecular line maps that can be compared directly with
observations (Padoan et al. 1998). CO spectra show typically several peaks
corresponding to different high density sheets of gas along the line of
sight. The average spectrum is almost Gaussian.

Padoan et al. (1999a) have compared the computed spectra in detail with CO
observations made of the Perseus molecular cloud complex. The statistical
properties of the computed spectra are almost identical to the properties of
the spectra observed e.g. in L\,1448. This demonstrates that the main
features of such clouds are correctly described by the MHD models.

The models spectra can be analyzed using the LTE approximation in order to
estimate the validity of the LTE assumption in the interpretation of
observations. Even in isothermal models one can find a wide range of
excitation conditions and LTE analysis tends to underestimate the true
column densities. In the case of our models the discrepancy can be as high
as a factor $\sim$5, at the lowest values of the CO column density. 
(Padoan et al. 1999b)

\section{Thermal equilibrium of interstellar clouds}

The previous model calculations can be extended to the study of thermal
balance in interstellar clouds. The MHD calculations provide the heating
rates due to ambipolar diffusion and total heating rates are obtained by
adding other known mechanisms (cosmic ray heating etc.). These are balanced
by cooling rates which in the case of molecular clouds are mainly due to
line emission.

The cooling rates at any point in the cloud will depend on the local escape
probability of the emitted photons but also on the photon flux from the
surrounding regions. Since velocity and density distributions are
inhomogeneous the calculation of the cooling rates requires proper solution
of the full three-dimensional radiative transfer problem.

In the case of Monte Carlo method cooling rates are a by-product of the
radiative transfer simulation and are obtained by simply counting the net
flow of photons from each cell. For practical reasons each molecule is
simulated separately and the kinetic temperatures, $T_{\rm kin}$, can be
updated only after all the main cooling species have been simulated. Since
the line emission depends in turn on T$_{\rm kin}$ the final solution is
obtained with iteration. Since $^{12}$CO is usually the dominant coolant it
is not necessary to update the cooling rate estimates from other species on
every iteration.

The ambipolar diffusion heating can exceed heating by cosmic rays (Padoan,
Zweibel \& Nordlund 1999c). Computed cloud temperatures are therefore
sensitive to initial assumptions and can be used to place limits on the
possible range of the input parameters, especially the strength of the
magnetic fields.

\end{document}